\documentclass[a4paper,10pt]{article}
\usepackage{amssymb}
\date{}
\title{ Scalar-Graviton Scattering in Noncommutative Space and Deformed Newton Gravity }
\author{ A. $\textrm{Jahan}^*$, A. $\textrm{Parvishi}^{**}$\\
\large Research Institute for Astronomy and Astrophysics of Maragha\\
\large (RIAAM) – Maragha, IRAN, P. O. Box: 55134 - 44$1^*$\\
\large Islamic Azad University, Urmia Branch, Urmia, $\textrm{Iran}^{**}$\\
\large jahan@riaam.ac.ir}
\begin{document}
\maketitle
\begin{abstract}
The isotropy of Newton potential in noncommutative space disappears as it deforms to a momentum dependent one. We generalize the earlier derivation of such a deformed potential to the relativistic regime by calculating the 2 scalar-1 graviton scattering amplitude by taking into account the noncommutativity of space.\\\
\end{abstract}

\section*{\large 1.\quad Introduction}
There are several serious motivations, arising from the different arenas of the planck scale physics [1-3], to consider a noncommutative (NC) algebra for the coordinates $x^\mu$ spaning the space-time manifold, via
\begin{equation}\label{1}
[\widehat{x}^\mu,\widehat{x}^\nu]=i\theta^{\mu\nu},
\end{equation}
where $\theta^{\mu\nu}$ assumed to be a constant antisymmetric matrix. As a result of "$\theta$-deformation" of the algebra of space-time coordinates one must replace the usual product among the fields with Weyl-Moyal product or $\star$-product [4], i.e.
\begin{equation}\label{1}
\phi_1(\widehat x)\star\phi_2(\widehat x)\equiv \lim_{x\rightarrow y}e^{\frac{i}{2}\theta^{\mu\nu}\partial^x_\mu\partial^y_\nu}\phi_1(x)\phi_2(y).
\end{equation}
From the Feynman's rules point of view the only effect of the $\star$-product is to modify the $n$-point interaction vertices (3$\leq n$) by the phase factor [4]
\begin{equation}\label{1}
\tau(p_1,\ldots,p_n)=e^{-\frac{i}{2}\sum^n_{a<b}\,p_a\wedge p_b},
\end{equation}
where $p_a\wedge p_b=\theta_{\mu\nu}p^\mu_a p^\nu_b$. Here the momentum flow of the $a$-th field into the vertex is denoted by $p_a$. In the case of noncommutative QED (QED living on a NC space-time), by analyzing the electron-photon interaction vertex one can demonstrate that at the quantum mechanical level the $\theta$-deformation of spatial coordinates gives rise to a deformed Coulomb potential [5]
\begin{equation}\label{1}
V_\theta=-\frac{Ze^2}{|\vec x|}-\frac{Ze^2}{4|\vec x|^3}\vec L\cdot\vec\theta+O(\theta^2),
\end{equation}
with $\theta_{ij}=\frac{1}{2}\epsilon_{ijk}\theta_k$ and $\vec L=\vec x\times\vec p$. However one may defer the field theoretic considerations and follow a more economical way to achieve (4) by redefining the coordinates
\begin{equation}\label{1}
\widehat x_i\rightarrow x_i-\frac{\theta_{ij}}{2}p_j.
\end{equation}
in the Coulomb potential $\widehat V=-\frac{Ze^2}{\sqrt{\widehat x_i\widehat x_i}}$ defined over a noncommutative space. The prescription which has been also followed to consider the $\theta$-deformation of the Newton potential and its possible classical phenomenological consequences [6, 7]. \\\
So, as the aim of this letter, it seems to be a logical step to deduce the deformed Newton potential by looking at the mater-graviton interaction vertex
in a NC background space. In the next section we write down the explicit form of the momentum space vertex factor for a Klein-Gordon field scattering from a graviton in NC space. In section 3, the two-body scattering problem in NC flat background is considered and the deformed Newton potential is re-derived by calculating the Fourier transform of the scattering amplitude. The scalar particle scattering off a static source of the graviton is also considered in section 4. Our result is the relativistic generalization of the previous calculations [6, 7]. In this work we assume $c=\hbar=1$ and $\eta_{\alpha\beta}=\textrm{diag}(1,-1,-1,-1)$.
\section*{\large 2.\quad Matter-Graviton Coupling}
The Klein-Gordon lagrangian density in curved space-times is given by
\begin{equation}\label{1}
\mathcal{L}_{KG}=\frac{\sqrt{-g}}{2}(\partial_\alpha{\phi}\partial^\alpha{\phi}-m^2\phi^2),
\end{equation}
By expanding the metric tensor and its determinants up to the first order in parameter $\kappa=\sqrt{32\pi G}$ as [8, 9, 10]
{\setlength\arraycolsep{2pt}
\begin{eqnarray}\label{1}
g_{\mu\nu}&=& \eta_{\mu\nu}+\kappa h_{\mu\nu}+O(\kappa^2), \\
g&=&1+\frac{\kappa}{2}h+O(\kappa^2),
\end{eqnarray}}
where $g=\det g_{\mu\nu}$ and $h=\eta^{\mu\nu}h_{\mu\nu}$, and then substituting them in (6) we find the Lagrangian density
\begin{equation}\label{1}
\mathcal{L}_{KG}=\frac{1}{2}(\partial_\alpha{\phi}\partial^\alpha{\phi}-m^2\phi^2)-\frac{\kappa}{2}{h}^{\mu\nu}
\Big[\partial_\mu{\phi}\partial_\nu{\phi}-\frac{1}{2}(\partial_\alpha{\phi}\partial^\alpha{\phi}-m^2\phi^2)\Big]+O(\kappa^2).
\end{equation}
Now the second term of (9), first order in $\kappa$,  clearly describes the 2 scalar-1 graviton interaction in a flat background. Thus we have the interaction term as
\begin{equation}\label{1}
\mathcal{L}_{int}=-\frac{\kappa}{2}{h}^{\mu\nu}
\Big[\partial_\mu{\phi}\partial_\nu{\phi}-\frac{1}{2}\big(\partial_\alpha{\phi}\partial^\alpha{\phi}-m^2\phi^2\big)\Big].
\end{equation}
from which one obtains the corresponding vertex factor [8, 9, 10]
\begin{equation}\label{1}
\tau_{\alpha\beta}(p,p')=-\frac{i\kappa}{2}(p_\alpha p'_\beta+p'_\alpha p_\beta-\eta_{\alpha\beta}p\cdot p')
\end{equation}
In de Donder gauge (harmonic gauge) $\partial^\alpha h_{\alpha\beta}-\frac{1}{2}\partial_\beta h=0$ the Einstein-Hilbert action takes the form
\begin{equation}\label{1}
S_{EH}=\frac{1}{2}\int d^4x h_{\mu\nu}Q^{\mu\nu,\alpha\beta}h_{\alpha\beta}
\end{equation}
with
\begin{equation}\label{1}
Q^{\mu\nu,\alpha\beta}=\frac{1}{2}(\eta^{\mu\alpha}\eta^{\nu\beta}+\eta^{\mu\beta}\eta^{\nu\alpha}-\eta^{\mu\nu}\eta^{\alpha\beta})\partial^2
\end{equation}
So the momentum-space graviton propagator is found to be
\begin{equation}\label{1}
D_{\mu\nu,\alpha\beta}(q)=-\frac{i}{2q^2}(\eta_{\mu\alpha}\eta_{\nu\beta}+\eta_{\mu\beta}\eta_{\nu\alpha}-\eta_{\mu\nu}\eta_{\alpha\beta})
\end{equation}

\section*{\large 3.\quad Two Body Scattering in NC Space}
In a NC flat background the interaction term must be changed to [11]
\begin{equation}\label{1}
\mathcal{L}_{int}=-\frac{\kappa}{2}{h}^{\mu\nu}\star
\Big[\partial_\mu{\phi}\star\partial_\nu{\phi}-\frac{1}{2}\big(\partial_\alpha{\phi}\star\partial^\alpha{\phi}-m^2\phi\star\phi\big)\Big].
\end{equation}
Therefore, with the aid of formula (3) we obtain the deformed momentum space 2 scalar-1 graviton vertex factor as
\begin{equation}\label{1}
\tau_{\alpha\beta}^\theta(p,p')=-\frac{i\kappa}{2}(p_\alpha p'_\beta+p'_\alpha p_\beta-\eta_{\alpha\beta}p\cdot p')e^{\frac{i}{2}\vec p\wedge
\vec p'}.
\end{equation}
where we have assumed $\theta_{\mu 0}=0$ that implies $\vec{p}\wedge \vec{q}\rightarrow\theta_{ij}p_i q_j$ to avoid the problematic features of the NC models [4]. Now let us look at a typical two-body scattering mediated by a graviton. For the scalar particles with masses $m_1$ and $m_2$ the scattering amplitude is
{\setlength\arraycolsep{2pt}
\begin{eqnarray}\label{8}
\mathcal M_\theta&=&\tau^{\mu\nu}_\theta(p_1,p'_1)D_{\mu\nu,\alpha\beta}(p_1-p'_1)\tau^{\alpha\beta}_\theta(p_2,p'_2)\\\nonumber
&=&\frac{4\pi G}{(p_1-p'_1)^2}\Bigg\{\Big[(p_1+p_2)^2-m^2_1-m^2_2\Big]^2+\Big[(p_1-p'_2)^2-m^2_1-m^2_2\Big]^2\\\nonumber
&-&\Big[(p'_1-p_1)^2+4m^2_1m^2_2\Big]^2\Bigg\}e^{i\vec p\wedge
(\vec p-\vec p')}
\end{eqnarray}}
In the non-relativistic limit we have
{\setlength\arraycolsep{2pt}
\begin{eqnarray}\label{8}
(p'_1-p_1)^2&\approx& -\vec q^{\,2}\\
(p_1+p_2)^2&\approx&(m_1+m_2)^2\\
(p_1-p'_2)^2&\approx&(m_1-m_2)^2+\vec q^{\,2}
\end{eqnarray}}
By substituting (18)-(20) in (17) we find the deformed gravitational potential
{\setlength\arraycolsep{2pt}
\begin{eqnarray}\label{8}
U_{\theta}(\vec x)&=&-\frac{1}{4m_1m_2}\int\frac{d^3q}{(2\pi)^3}e^{i\vec x\cdot\vec q}\mathcal M_{\theta}(\vec q\,)\\\nonumber
&=&-G\frac{m_1m_2}{\sqrt {({x_i-\frac{1}{2}\theta_{ij}p_j})({x_i-\frac{1}{2}\theta_{ik}p_k })}}+4\pi G\delta(\vec{\widetilde{x}}\,)
\end{eqnarray}}
where
\begin{equation}\label{1}
\widetilde{x}_i=x_i-\frac{\theta_{ij}}{2}p_j
\end{equation}
One must note that the second term of (21) is repulsive and can only be measured for bound s-states and thus does not appear for gravitational field as a spin zero massless field [10].

\section*{\large 4.\quad Scattering off a Static Graviton Source }
The linearized Einstein equation in de Donder gauge reads
\begin{equation}\label{8}
\Box\Big[h_{\mu\nu}(x)-\frac{1}{2}\eta_{\mu\nu}h(x)\Big]=-\kappa T_{\mu\nu}(x).
\end{equation}
On the other hand for a static source of graviton with mass $M$ we have the energy-momentum tensor as $T_{\mu\nu}(x)=M\delta_{\mu 0}\delta_{\nu 0}\delta(\vec{x})$. Hence in momentum space one finds
\begin{equation}\label{8}
h_{\mu\nu}(q)=\frac{\kappa M}{4\vec {q}^{\,2}}(\eta_{\mu\nu}-2\eta_{\mu 0}\eta_{\nu 0})2\pi\delta(q_0).
\end{equation}
Therefore, having at hand the ingredients needed for calculation of the scattering amplitude $\mathcal M$ we get [10]
{\setlength\arraycolsep{2pt}
\begin{eqnarray}\label{8}
i\mathcal M_\theta&=&ih_{\mu\nu}(\vec{q}\,)\tau^{\mu\nu}_\theta(p,q),\\\nonumber
&=&-\frac{4\pi GM}{\vec {q}^{\,2}}(2m^2+4\vec {p}^{\,2})e^{\frac{i}{2}\vec p\wedge\vec q}.
\end{eqnarray}}
The Fourier transform of scattering amplitude leaves us with the Newton potential as
{\setlength\arraycolsep{2pt}
\begin{eqnarray}\label{1}
U_\theta&=&-4\pi GM\frac{m^2+2\vec {p}^{\,2}}{E}\int\frac{d^3q}{(2\pi)^3}\frac{e^{i\vec x\cdot\vec q+\frac{i}{2}\theta_{ij}p_i q_j}}{\vec {q}^{\,2}},\\\nonumber
&=&-\frac{m^2+2\vec {p}^{\,2}}{E}\,\frac{GM}{\sqrt {({x_i-\frac{1}{2}\theta_{ij}p_j})({x_i-\frac{1}{2}\theta_{ik}p_k })}},\\\nonumber
&=&-\frac{m^2+2\vec {p}^{\,2}}{E}\Bigg(\frac{GM}{|\vec x|}+\frac{GM}{4|\vec x|^3}\vec L\cdot\vec\theta+O(\theta^2)\Bigg).
\end{eqnarray}}
In non-relativistic limit, i.e. $|\vec p|<<m$ and $E\approx m$ the above result coincides with that of [6, 7]
{\setlength\arraycolsep{2pt}
\begin{eqnarray}\label{1}
U^{non-rel}_\theta=-\frac{GM m}{|\vec x|}-\frac{GM m}{4|\vec x|^3}\vec L\cdot\vec\theta+O(\theta^2)
\end{eqnarray}}
which was gained by implementing (5) in $\widehat U=-G\frac{mM}{\sqrt{\widehat x_i\widehat x_i}}$. For a relativistic particle with $E\simeq |\vec p|$ we have
{\setlength\arraycolsep{2pt}
\begin{eqnarray}\label{1}
U^{rel}_\theta&=&-\frac{2GME}{|\vec x|}-\frac{GME}{2|\vec x|^3}\vec L\cdot\vec\theta+O(\theta^2).
\end{eqnarray}}
where the first term of (17) is the well-known result arising from the scalar-graviton scattering in absence of the noncommutativity [10].
\section*{\large 5.\quad Conclusion}
We followed a field theoretic approach to re-derived the $\theta$-deformed Newton potential by calculating the matter-graviton scattering amplitude in a NC flat background. In the case of static graviton source our result confirms the earlier derivation and generalizes it to the relativistic regime.
\section*{\large Acknowledgments}
This work has been supported financially by Research Institute for Astronomy $\&$ Astrophysics of Maragha.
\section*{\large References}
[1]\hspace{0.2cm}N. Seiberg, E. Witten, JHEP \textbf{9909} (1999) 032.\\\
[2]\hspace{0.2cm}H. S. Snyder, Phys. Rev. \textbf{71} (1947) 38.\\\
[3]\hspace{0.2cm}S. Doplicher, K. Fredenhagen, and J. E. Roberts, Commun. Math. Phys. \textbf{172} (1995) 187.\\\
[4]\hspace{0.2cm}R. Szabo, Phys. Rept. \textbf{378} (2003) 207.\\\
[5]\hspace{0.2cm}M. Chaichian, M. M. Sheikh-Jabbari, and A. Tureanu, Phys. Rev. Lett. \textbf{86} (2001) 2716.\\\
[6]\hspace{0.2cm}J M. Romero, J. A. Santiago, J. D. Vergara, Phys.Lett. \textbf{A310} (2003) 9.\\\
[7]\hspace{0.2cm} B. Mirza, M. Dehghani, Commun. Theor. Phys. \textbf{42} (2004) 183.\\\
[8]\hspace{0.2cm}N. E. J. Bjerrum-Bohr, Ph.D. thesis, The Niels Bohr Institute, University of Copenhagen, July 2004.\\\
[9]\hspace{0.2cm}B. Holstein, Am. J. Phys. \textbf{74} (2006) 1002.\\\
[10]\hspace{0.2cm}M. Scadron, \emph{Advanced Quantum Theory and its Applications
through Feynman Diagrams}, Springer-Verlag, New York (1979).\\\
[11]\hspace{0.2cm}J. W. Moffat, Phys. Lett. \textbf{B493} (2000) 142.\\\

\end{document}